\documentclass[12pt]{article}

\newenvironment{wileykeywords}{\textsf{Keywords:}\hspace{\stretch{1}}}{\hspace{\stretch{1}}\rule{1ex}{1ex}}

\usepackage{amsmath,amssymb}
\usepackage{graphicx}
\usepackage{color}
\usepackage{dcolumn}
\usepackage{bm}
\usepackage[numbers,super,comma,sort&compress]{natbib}
\usepackage{hyperref}
\usepackage[font=scriptsize,labelfont=bf]{caption}

\definecolor{background-color}{gray}{0.98}

\title{iGVPT2 : an interface to computational chemistry packages for anharmonic corrections to vibrational frequencies}

\author{
Lo\"{i}c Barnes,
Baptiste Schindler,
Isabelle Compagnon,
Abdul-Rahman Allouche\thanks{ Univ Lyon, Universit\'e Claude Bernard Lyon 1, CNRS, Institut Lumi\`ere Mati\`ere, F-69622, VILLEURBANNE, France}
}

\usepackage{algorithm}
\usepackage[noend]{algpseudocode}
\newcommand{\algoHarmonic}{
\begin{algorithm}[H]
\scriptsize
\begin{algorithmic}[1]
\State Set the High level method and the software to be used for calculating harmonic modes
\State Set the stepsize ($\delta$) to be used for calculating numerically the second derivatives along atom coordinates
\State Generate the input file to calculate the energy for the already optimized geometry.
\For{each atom $i$ in (number of atoms) }
\State Generate geometries along $x+\delta$, $x-\delta$, $y+\delta$, $y-\delta$, $z+\delta$ and $z-\delta$
\EndFor
\For{each geometry} \Comment{This loop can be run in parallel}
\State \hspace{\algorithmicindent} Compute energy and dipole (or gradients, recommended if the analytical calculation of derivatives is implemented in the computational chemistry package)
\EndFor
\State Using the calculated energies (or gradients if available), compute second derivatives of energy and first derivatives of dipole.
\State Using these derivatives, compute harmonic modes and (if required) the IR intensities at double harmonic approach
\end{algorithmic}
\caption{Harmonic pseudo-code}
\label{figure:algoHarmonic}
\end{algorithm}
}

\newcommand{\algoAnharmonic}{
\begin{algorithm}[H]
\scriptsize
\begin{algorithmic}[1]
\State Set the High level method and the software to be used
\State Optimized geometry using the high level method
\State Compute harmonic modes and IR intensities using the high level method
\State Set the low level method(\emph{ab initio}, DFT, AM1, DFTB3, MMFF94, ...) and the software to be used (Orca, Gaussian, FireFly, OpenMopac, DFTB+, ...). It is obviously possible to use the same method and the same software at high and low levels.
\State Generate the input file, for the low level method using the optimized geometry at High level method 
\State Select the modes to be used and set the value of $f$ to the number of these modes
\State Generate all geometries needed to calculate the cubic and quartic derivatives at the low level method
\For{each generated file} \Comment{\textbf{This loop can be run in parallel}}
\State \hspace{\algorithmicindent} Compute energy and dipole
\EndFor
\State Using the calculated energies, compute cubic and quartic energy derivatives
\State Using the calculated dipoles, compute second and cubic dipole derivatives if IR is required
\State Using these derivatives, compute anharmonic frequencies and (if required) the IR intensities using VPT2, DCPT2, HDCP2, VPT2+K approaches.
\end{algorithmic}
    \caption{Anharmonic VPT2 pseudo-code}
    \label{figure:algoAnharmonic}
\end{algorithm}
}

\begin{document}

\maketitle

\begin{abstract}
iGVPT2 is a program for computing anharmonic corrections to vibration frequencies, based on force field expansion of the potential energy surface in normal mode coordinates. 
It includes second order vibrational perturbation theory (VPT2) algorithm and its derived methods (VPT2+K, DCPT2, HDCPT2). iGVPT2 is interfaced with several computation chemistry packages to compute the potential energies and dipoles derivatives. The second, third and quartic derivatives can be computed at the same level of theory but they can be also computed using different methods via one or two computational packages. iGPVT2 includes also a very fast hybrid QM//MM approach for biomolecules. It is provided free-of-charge for non-commercial research (see \url{https://sites.google.com/site/allouchear/igvpt2}).
\end{abstract}

\begin{wileykeywords}
Anharmonic, VPT2, VPT2+K, GVPT2, DCPT2, HDCPT2
\end{wileykeywords}

\clearpage

\section*{\sffamily \Large Introduction} 

The comparison between measured infrared spectra (IR) and calculated ones is an example of successful application of theoretical chemistry for generating fingerprints of molecular structure. The standard approach used to compute IR spectra consists of quantum mechanics calculations in the static double harmonic approximation as implemented in most major computational chemistry packages (CCP). However, in this approach, the anharmonic effects on the fundamental modes are not taken into account and band combinations and overtones cannot be calculated. One or more scaling factors are often applied to the computed harmonic frequencies to reduce the difference with experimental values.
Several methods are proposed to include explicitly anharmonic effects in quantum chemistry vibrational frequency calculations however, two of them being widely used. Firstly, the vibrational self-consistent field (VSCF)\cite{Bowman78} followed by a second order perturbation corrected theory (referred as VSCF-PT2\cite{Yagi00} or as cc-VSCF\cite{Jung96} in the literature) for weak coupling between modes or by a vibrational interaction configuration (VCI)\cite{Christoffel82} in the case of very strong coupling. This method is implemented in several CCP (Gamess-US\cite{gamess}, Molpro\cite{molpro}, NWChem\cite{nwchem}). The second most used method is the vibrational second-order perturbation theory (VPT2) approach. It uses quadratic, all relevant cubic, and quartic force constants to create a quartic force field (QFF) in order to calculate anharmonic vibrational frequencies of polyatomic molecules. The derivatives are calculated with an \emph{ab initio} potential. This method is implemented in two CCP packages: CFOUR\cite{CFOUR} and Gaussian\cite{G09}. The critical point for the calculation of energies in this approach is the Fermi resonance which may result in unphysical anharmonic corrections. Martin et al.\cite{Martin95} proposed a procedure to identify and remove resonant terms in a first step and use a variational treatment in a second step. This procedure is referred to as VPT2+K\cite{Rosnik14} by some authors and by GVPT2 (generalized VPT2) by others.\cite{Barone14} 
As a full variational treatment would become prohibitive for large molecules, a non variational called degeneracy-corrected VPT2 (DCPT2)\cite{Kuhler96} is proposed in the literature.
The use of a transition function between the standard VPT2 formulation far from resonance and the DCPT2, called Hybrid DCPT2 (HDCPT2)\cite{Bloino12} was proposed to reduce the error obtained with DCPT2.  All these methods have been implemented in the Gaussian package. 
In this paper we describe a new program (called iGVPT2) to compute the infrared spectra including anharmonic effects. Compared to other available softwares, the main advantages of iGVPT2 are:
\begin{itemize}
\item{iGVPT2 is interfaced with several computational chemistry packages. In the present version, it supports Orca\cite{orca}, Gaussian\cite{G09}, Molpro\cite{molpro}, OpenMopac\cite{openmopac}, FireFly\cite{gamess,firefly}, Gamess-US\cite{gamess}, DFTB+\cite{Aradi07}, but can be easily extended to others CCP. Compared to others implementation of (G)VPT2, iGVPT2 can be used with several CCP while VPT2 implemented in CFOUR\cite{CFOUR} is usable only with its \emph{ab initio} methods (This for instance does not include DFT, which is not implemented in CFOUR) and GVPT2 implemented in Gaussian\cite{G09} is usable with the computational methods implemented in it.}
\item{Any hybrid calculation using a high level method for harmonic part and low level method for cubic and quartic derivatives can be done using iGVPT2. For example, the user can compute an anharmonic infrared spectrum using B2PLYP in Orca\cite{orca} to calculate the harmonic part and the OpenMopac\cite{openmopac} to calculate the cubic and quartic derivatives. This approach is similar to that proposed in Refs. \cite{Brauer04,Roy13} for VSCF calculations, applied here for GVPT2 calculations.}
\item{Calculations of cubic and quartic derivatives can be done in parallel and the real time calculation can be divided by a factor of about $\sim f^3$, where f is the number of harmonic frequencies.}
\item{A hybrid \emph{ab initio}//MMFF94\cite{Barnes16} is implemented in iGVPT2. It is a very fast and accurate approach to compute anharmonic spectra for large biomolecules.}
\item{For large molecules, a highly parallelized method is implemented in iGVPT2 for numerical calculation of harmonic modes. The real time calculation is $6N+1$(where $N$ is number of atoms) smaller than the cputime.}
\end{itemize}

We describe below the methods implemented in iGVPT2. It is followed by examples of results obtained with our code and compared to these obtained with other softwares for 3 molecules.  Finally, we provide the conclusions and outlook for our software.

\section*{\sffamily \Large Methodology}

An anharmonic calculation using VPT2 approach, or its variants, is performed in 3 steps. It starts by the harmonic frequencies and normal mode calculations. For small and medium-sized molecules, it is recommended to use, in this step one of the available CPP where analytical second derivatives are implemented. For larger system (typically more than 30 atoms), iGVPT2 can compute numerically the harmonic modes. This part of iGVPT2 is parallelized. This approach is shown in Algorithm \ref{figure:algoHarmonic}. 
\footnotesize{
    \algoHarmonic
}

In the second step of the anharmonic calculation, the cubic and fourth derivatives must be calculated. The potential energy surface is approximated by a quartic force field (QFF). It is given by,
\begin{flalign*}
&V^{QFF}(\bold{Q}) = V_0 + V_1(\bold{Q}) + V_2(\bold{Q}) + V_3(\bold{Q})\\
&V_1(\bold{Q})     = \sum \limits_{i=1}^{f} { \frac{1}{2} h_i Q_i^2 + \frac{1}{6} t_{iii} Q_i^3 + \frac{1}{24} u_{iiii} Q_i^4}\\
&V_2(\bold{Q})     = \sum \limits_{ij, i \neq j}^{f} {\frac{1}{2} t_{ijj} Q_i Q_j^2 + \frac{1}{6} u_{ijjj} Q_i Q_j^3} + \sum \limits_{ij,i<j}^{f} {\frac{1}{4} u_{iijj} Q_i^2 Q_j^2} \\
&V_3(\bold{Q})     = \sum \limits_{ijk, i \neq j<k}^{f} { t_{ijk} Q_i Q_j Q_k} + \sum \limits_{ijk,i \neq j<k}^{f} {\frac{1}{2} u_{iijk} Q_i^2 Q_j Q_k}
\end{flalign*}

Where $f$, $\bold{Q}$, $V_0$, $h_i$, $t_{ijk}$ and $u_{ijkl}$ denote the number of normal modes, normal coordinates, the energy and its second-, third-, and fourth-order derivatives with respect to the normal coordinates at the equilibrium geometry, respectively.
The derivatives are calculated through numerical differentiations of the energy using any \emph{ab initio} or (semi-)empirical method implemented in any quantum chemistry software. 
The QFF can be built at 1-mode (1MR), 2-modes (2MR) or 3-modes coupling (3MR) representation. To compute all terms in the QFF potential, $1+2f$, $1+6f^2$ and $1+6f^2 + 8f(f-1)(f-2)/6$ single point calculations are required for 1MR, 2MR and 3MR respectively.
iGVPT2 can compute these energies and then the derivatives at any level of theory using any CCP. Alternatively, these derivatives can be calculated by iGVPT2 for the MMFF94\cite{Halgren96} potential. In this case, iGVPT2 uses OpenBabel\cite{openbabel} library to make this step. 
The derivatives are calculated by numerical differentiation of the energy, using the formula given by Yagi et al.\cite{Yagi04} excepted for the $t_{iii}$ and $u_{iiii}$ 1-mode coupling terms and for the 2-modes $u_{iijj}$ term where we used the following formula:
\begin{flalign*}
&t_{iii} = \frac{1}{8 \delta_{i}^{3}} [ -V(+3\delta_{i})  + 8 V(+2\delta_{i})  -13 V(+\delta_{i}) +13 V(-\delta_{i}) -8V(-2\delta_{i}) +  V(-3\delta_{i}) ]\\
&u_{iiii} = \frac{1}{6 \delta_{i}^{4}} [-V(+3\delta_{i})  + 12 V(+2\delta_{i})  -39 V(+\delta_{i}) + 56 V_0 -39 V(-\delta_{i}) +12V(-2\delta_{i}) -  V(-3\delta_{i}) ]\\
&u_{iijj} = \frac{1}{\delta_{i}^{2}\delta_{j}^{2}}\{[V(+\delta_{i},+\delta_{j}) + V(-\delta_{i},+\delta_{j}) + V(+\delta_{i},-\delta_{j}) +V(-\delta_{i},-\delta_{j})]\\
&\qquad\qquad\quad-2[V(\delta_{i})  + V(-\delta_{i}) + V(\delta_{j})  + V(-\delta_{j})] + 4 V_0\}
\end{flalign*}
where $\delta_{i}$ denotes the stepsize for the i$^{th}$ normal coordinate, and $V(n \delta_{i})$ denotes the energy at $Q(n \delta_{i})$ and fixing to zero the other coordinates.

To compute the infrared intensities, the first and second dipole derivatives are also calculated using the formula given by,
\begin{flalign*}
&\mu^{x}_{i} = \frac{1}{60 \delta_{i}} [ \mu^{x}(+3\delta_{i})  - 9 \mu^{x}(+2\delta_{i})  +45 \mu^{x}(+\delta_{i}) -45 \mu^{x}(-\delta_{i}) +9 \mu^{x}(-2\delta_{i}) -  \mu^{x}(-3\delta_{i}) ]\\
&\mu^{x}_{ii} = \frac{1}{180 \delta_{i}^{2}} [2 \mu^{x}(+3\delta_{i})  -27 \mu^{x}(+2\delta_{i})  +270 \mu^{x}(+\delta_{i}) -490 \mu^{x}_0 +270 \mu^{x}(-\delta_{i}) -27 \mu^{x}(-2\delta_{i}) + 2 \mu^{x}(-3\delta_{i}) ]\\
&\mu^{x}_{ij} = \frac{1}{4 \delta_{i} \delta_{j}} [+\mu^{x}(+\delta_{i}, +\delta_{j})  -  \mu^{x}(+\delta_{i},-\delta_{j})  - \mu^{x}(-\delta_{i}, +\delta_{j}) +  \mu^{x}(-\delta_{i},-\delta_{j}) ]
\end{flalign*}
where $\mu^{x}(n \delta_{i}, m \delta_{j})$ denotes the $x$ component of dipole at $Q(n \delta_{i}) + Q(m \delta_{j})$ when other coordinates are fixed to zero. Similar formula were used for $y$ and $z$ components. Note that, if the first derivatives of the dipole are calculated by the CCP at the high level method, these derivatives can then be used to compute the anharmonic infrared spectrum instead of the first derivatives calculated numerically using the low level method of computation chemistry method. The third derivatives are calculated using the formula used for energy derivatives.

Finally, the anharmonic fundamental frequencies are calculated using either VPT2 (without any treatment of resonances), GVPT2 (Fermi resonances are treated variationally), or non variational approaches such as DCPT2 or HDCPT2 methods.  For GVPT2 calculation, we used the formula given in Ref.\cite{Rosnik14}, Fermi resonances are treated using the Martin et al.\cite{Martin95} criteria. The intensities are calculated using the formula given in Ref.\cite{Barone14} Our approach is illustrated in Algorithm \ref{figure:algoAnharmonic} and in Figure \ref{figure:VPT2Scheme}

\begin{center}
\setlength{\fboxsep}{0pt}%
\setlength\fboxrule{0pt}
\fbox{
\begin{minipage}{0.5\textwidth}
\vtop{
\footnotesize{
    \algoAnharmonic
}
}
\end{minipage}
\begin{minipage}{0.5\textwidth}
\centering
	\includegraphics[scale=1.75]{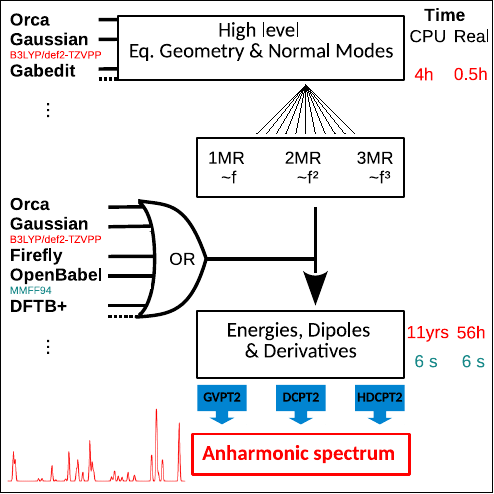}
	\captionof{figure}{Anharmonic VPT2 scheme}
	\label{figure:VPT2Scheme}
\end{minipage}
}
\end{center}

\section*{\sffamily \Large Program and supported computational chemistry packages}

iGVPT2 is written in C using two libraries: cchemilib (written in C and developed by one of the authors of this paper) and OpenBabel\cite{openbabel} written in C++.
At least one computation chemistry package is required to use iGVPT2. In the first step of an anharmonic calculation with iGVPT2, the harmonic modes must be calculated by a CCP.
If the harmonic calculation is not implemented in the CCP, iGVPT2 can compute these frequencies using the energies (or the gradients if available in the CCP).
Presently, iGVPT2 can read harmonic modes from Orca, Gaussian, Molpro, OpenMopac, FireFly, Gamess-US output files and from a Gabedit\cite{gabedit} file. iGVPT2 can also compute the harmonic frequencies using the energies computed by Orca, Gaussian, Molpro, OpenMopac, FireFly, Gamess-US and DFTB+ or other CCP using a plugin to be written by the user (see later). 
After reading (or calculating) harmonic frequencies, iGVPT2 generates input files for a single energy calculation for CCP (to be chosen by the user) supported by iGVPT2. 
iGVPT2 runs then the CCP for each file and retrieves the calculated energies needed to compute third and fourth derivatives. For this step, Orca, Gaussian, Gamess-US, Molpro, OpenMopac, FireFly and DFTB+  are supported by iGVPT2. Other CCP can be supported via a plugin. If MMFF94 is chosen by the user in this step, no file is generated. iGVPT2 uses the openbabel library to compute needed energies.
Using calculated energies (by CCP or by iGVPT2 if MMFF94 is selected),  iGVPT2 computes the derivatives, the anharmonic frequencies and the infrared intensities.
As mentioned above, iGVPT2 supports already several CCP but it can support any other CCP if the user writes a specific plugin for his favorite CCP. In this case and for each calculation, iGVPT2 generates an ascii file containing an integer to specify the type of calculation (energy of gradients), the charge of molecule, the multiplicity and the geometry. The plugin (to be written by the user) must read this file, 
create an input file for his favorite CCP, run it, read energy, dipole (and gradients if required), and save it in a new ascii file. iGVPT2 reads finally the needed values from this file.
In fact iGVPT2 supports Gamess-US, Molpro and DFTB+ via plugins. The plugin of DFTB+ is writen in C++ and that of Gamess-US and Molpro are bash scripts. The code sources for these 3 plugins are distributed with iGVPT2 and can be used to write other plugin for other CCP.

\section*{\sffamily \Large Results}

\subsection*{\sffamily \large Validation}
To validate our implementation, we calculated the fundamental frequencies of three molecules: Water, C$_2$H$_4$ and H$_2$CO using three methods implemented in iGVPT2 namely GVPT2, DCPT2 and HDCPT2.
These values are calculated using Gaussian\cite{G09} software and ours. The difference between our implementation and that of Gaussian\cite{G09} is the method used to compute the third and forth derivatives.
In Gaussian the derivatives are calculated by numerical differentiation of second derivatives that are computed analytically. By default, a stepsize of $0.01 \AA$ is used for all modes. 
In our implementation the derivatives are calculated numerically using the dimensionless reduced coordinate proposed by Yagi et al.\cite{Yagi04}, defined by 

$$ y_i = \sqrt{\frac{\omega_i}{\hbar}} Q_i$$
where $\omega_{i}$ and $\hbar$ denote the angular frequency for the i$^{th}$ normal mode and the Planck constant divided by $2 \pi$, respectively.

In this approach, by using an identical stepsize ($\delta y_i$) for all the reduced coordinates ($y_i$), different stepsizes  $\delta_i$ are generated for each mode, thus allowing larger and smaller stepsizes for lower and higher frequency modes, respectively. As recommended by Yagi et al.\cite{Yagi04} 0.5 was employed as a default value for the stepsize of reduced coordinates in our code.
Table \ref{table:compGaussian} shows the results using the default stepsize values in Gaussian and in iGVPT2. It is clear that the max deviation, for the three molecules and for the three methods, between the values obtained with the two software is about 4.5 $cm^{-1}$. This deviation is certainly due to the fact that the method for calculating the derivatives is not the same in the two softwares.

To further study the effect of the stepsize, we calculated the anharmonic GVPT2 values using different stepsizes with iGVPT2. 
Table \ref{table:reducedCoordinates} gives the fundamental frequencies of the three molecules studied here, obtained with iGVPT2 using GVPT2 approach, with different stepsizes of reduced coordinates for numerical differentiations. It shows that the dependence of the frequencies on the stepsize is small for values ranging from $0.3$ to $0.7$. 
Compared to the frequencies calculated with the stepsize 0.5, the max deviation is smaller than 7 cm$^{-1}$. 
In addition to the  approach based on reduced coordinates, we have implemented in iGVPT2 the possibility to compute numerically the  derivatives using the same stepsizes for all modes. Table \ref{table:normalCoordinates} gives the calculated frequencies using stepsizes ranging from $0.005 \AA$ to $0.075 \AA$. This table shows clearly a significant dependence on the value of stepsize. In conclusion, it is recommended to use the reduced coordinates with the default value (0.5) of the stepsize. For rigid molecules, a stepsize with smaller value could be used.

\begin{table}[H]
\scriptsize{
\begin{center}
\begin{tabular}{|l|l|l|l|ll|ll|ll|}

\hline
	&	\multicolumn{3}{|c|}{{Gaussian}} 		        &	\multicolumn{6}{|c|}{{iGVPT2/Orca}}								\\
\hline
	&	GVPT2	&	DCPT2	&	HDCPT2	&	\multicolumn{2}{|c|}{{GVPT2}}				&	\multicolumn{2}{|c|}{{DCPT2}}				&	\multicolumn{2}{|c|}{{HDCPT2}}				\\
\hline
H$_2$O 	&	1581.4	&	1581.1	&	1581.1	&	1581.9	&	(	0.5	)	&	1581.6	&	(	0.5	)	&	1581.6	&	(	0.5	)	\\
	&	3679.4	&	3680.1	&	3679.0	&	3678.9	&	(	-0.4	)	&	3679.7	&	(	-0.4	)	&	3678.5	&	(	-0.5	)	\\
	&	3776.3	&	3777.2	&	3776.3	&	3771.9	&	(	-4.4	)	&	3772.8	&	(	-4.4	)	&	3771.9	&	(	-4.4	)	\\
H$_2$CO	&	1173.2	&	1172.8	&	1172.8	&	1173.3	&	(	0.2	)	&	1172.9	&	(	0.1	)	&	1172.9	&	(	0.1	)	\\
	&	1252.6	&	1252.4	&	1252.4	&	1252.8	&	(	0.1	)	&	1252.5	&	(	0.1	)	&	1252.5	&	(	0.1	)	\\
	&	1518.3	&	1518.4	&	1518.4	&	1518.1	&	(	-0.2	)	&	1518.2	&	(	-0.2	)	&	1518.2	&	(	-0.2	)	\\
	&	1732.2	&	1732.2	&	1732.2	&	1732.3	&	(	0.1	)	&	1732.3	&	(	0.1	)	&	1732.3	&	(	0.1	)	\\
	&	2826.6	&	2826.8	&	2825.8	&	2824.6	&	(	-1.9	)	&	2824.8	&	(	-1.9	)	&	2823.8	&	(	-1.9	)	\\
	&	2856.6	&	2935.3	&	2934.3	&	2853.4	&	(	-3.1	)	&	2931.8	&	(	-3.5	)	&	2930.8	&	(	-3.5	)	\\
C$_2$H$_4$	&	823.1	&	821.6	&	821.6	&	822.7	&	(	-0.3	)	&	821.2	&	(	-0.4	)	&	821.2	&	(	-0.4	)	\\
	&	951.9	&	951.8	&	951.8	&	947.7	&	(	-4.2	)	&	947.7	&	(	-4.2	)	&	947.7	&	(	-4.2	)	\\
	&	955.9	&	955.8	&	955.8	&	955.1	&	(	-0.8	)	&	955.0	&	(	-0.8	)	&	955.0	&	(	-0.8	)	\\
	&	1044.2	&	1044.2	&	1044.2	&	1043.1	&	(	-1.1	)	&	1043.1	&	(	-1.1	)	&	1043.1	&	(	-1.1	)	\\
	&	1227.2	&	1227.2	&	1227.2	&	1225.8	&	(	-1.4	)	&	1225.8	&	(	-1.4	)	&	1225.8	&	(	-1.4	)	\\
	&	1357.5	&	1357.5	&	1357.5	&	1356.9	&	(	-0.6	)	&	1356.9	&	(	-0.6	)	&	1356.9	&	(	-0.6	)	\\
	&	1444.2	&	1444.0	&	1444.0	&	1444.8	&	(	0.5	)	&	1444.5	&	(	0.5	)	&	1444.5	&	(	0.5	)	\\
	&	1631.0	&	1643.6	&	1643.6	&	1630.4	&	(	-0.6	)	&	1643.0	&	(	-0.6	)	&	1643.0	&	(	-0.6	)	\\
	&	3011.2	&	3083.2	&	3082.6	&	3010.8	&	(	-0.3	)	&	3081.9	&	(	-1.3	)	&	3081.3	&	(	-1.3	)	\\
	&	3060.0	&	3060.4	&	3059.8	&	3058.8	&	(	-1.2	)	&	3059.2	&	(	-1.2	)	&	3058.6	&	(	-1.2	)	\\
	&	3126.0	&	3125.8	&	3125.2	&	3123.3	&	(	-2.7	)	&	3123.2	&	(	-2.6	)	&	3122.5	&	(	-2.7	)	\\
	&	3148.9	&	3149.0	&	3148.3	&	3146.3	&	(	-2.7	)	&	3146.4	&	(	-2.6	)	&	3145.7	&	(	-2.6	)	\\
	&		&		&		&		&				&		&				&		&				\\
MAX	&		&		&		&		&		4.4		&		&		4.4		&		&		4.4		\\
MAD	&		&		&		&		&		1.3		&		&		1.4		&		&		1.4		\\
RMSD	&		&		&		&		&		1.8		&		&		1.9		&		&		1.9		\\

\hline

\end{tabular}
\end{center}
} 
\caption{Fundamental anharmonic frequencies computed for three molecules using Gaussian software and using iGVPT2. For Gaussian calculations, the Coriolis contributions were neglected. In parentheses, the deviation between iGVPT2 and Gaussian values. The max deviation (MAX), root-mean-square deviation (RMSD) and average unsigned deviation (MAD) values are given in the last rows. All values are given in cm$^{-1}$}
\label{table:compGaussian}
\end{table}

\begin{table}[H]

\begin{center}
\scriptsize{
\setlength{\tabcolsep}{3pt}
\begin{tabular}{|r|rr|rr|rr|rr|rr|rr|rr|r|}

\hline
Coord.	&	\multicolumn{2}{|c|}{{0.1}}	&	\multicolumn{2}{|c|}{{0.2}}	&	\multicolumn{2}{|c|}{{0.3}}	&	\multicolumn{2}{|c|}{{0.4}}	&	\multicolumn{2}{|c|}{{0.5}}	&	\multicolumn{2}{|c|}{{0.7}}	&	\multicolumn{2}{|c|}{{0.9}}			\\
\hline
H$_2$O	&	1581.5	&	(	-0.4	)	&	1581.6	&	(	-0.3	)	&	1581.7	&	(	-0.2	)	&	1581.8	&	(	-0.1	)	&	1581.9	&	(	0.0	)	&	1582.2	&	(	0.3	)	&	1582.7	&	(	0.8	)	\\
	&	3680.1	&	(	1.1	)	&	3679.9	&	(	1.0	)	&	3679.7	&	(	0.8	)	&	3679.3	&	(	0.4	)	&	3678.9	&	(	0.0	)	&	3678.0	&	(	-0.9	)	&	3676.8	&	(	-2.1	)	\\
	&	3776.9	&	(	5.0	)	&	3776.3	&	(	4.4	)	&	3775.3	&	(	3.4	)	&	3773.8	&	(	1.9	)	&	3771.9	&	(	0.0	)	&	3766.7	&	(	-5.2	)	&	3759.6	&	(	-12.3	)	\\
H$_2$CO	&	1173.0	&	(	-0.3	)	&	1173.1	&	(	-0.2	)	&	1173.1	&	(	-0.2	)	&	1173.2	&	(	-0.1	)	&	1173.3	&	(	0.0	)	&	1173.4	&	(	0.1	)	&	1173.5	&	(	0.2	)	\\
	&	1252.8	&	(	0.1	)	&	1252.8	&	(	0.1	)	&	1252.8	&	(	0.1	)	&	1252.8	&	(	0.0	)	&	1252.8	&	(	0.0	)	&	1252.6	&	(	-0.1	)	&	1252.5	&	(	-0.3	)	\\
	&	1518.2	&	(	0.0	)	&	1518.2	&	(	0.0	)	&	1518.1	&	(	-0.1	)	&	1518.2	&	(	0.1	)	&	1518.1	&	(	0.0	)	&	1518.0	&	(	-0.2	)	&	1517.9	&	(	-0.3	)	\\
	&	1731.9	&	(	-0.4	)	&	1731.8	&	(	-0.5	)	&	1732.3	&	(	0.0	)	&	1732.3	&	(	0.0	)	&	1732.3	&	(	0.0	)	&	1731.9	&	(	-0.4	)	&	1731.9	&	(	-0.4	)	\\
	&	2825.2	&	(	0.6	)	&	2825.3	&	(	0.7	)	&	2825.2	&	(	0.5	)	&	2824.8	&	(	0.2	)	&	2824.6	&	(	0.0	)	&	2824.1	&	(	-0.6	)	&	2823.1	&	(	-1.5	)	\\
	&	2856.3	&	(	2.9	)	&	2856.0	&	(	2.6	)	&	2855.5	&	(	2.0	)	&	2854.6	&	(	1.1	)	&	2853.4	&	(	0.0	)	&	2850.3	&	(	-3.1	)	&	2846.1	&	(	-7.4	)	\\
C$_2$H$_4$	&	823.2	&	(	0.5	)	&	822.9	&	(	0.2	)	&	822.8	&	(	0.1	)	&	822.6	&	(	-0.1	)	&	822.7	&	(	0.0	)	&	822.5	&	(	-0.2	)	&	822.1	&	(	-0.6	)	\\
	&	923.1	&	(	-24.6	)	&	955.4	&	(	7.7	)	&	954.3	&	(	6.6	)	&	950.4	&	(	2.7	)	&	947.7	&	(	0.0	)	&	942.1	&	(	-5.6	)	&	936.8	&	(	-10.9	)	\\
	&	955.5	&	(	0.4	)	&	957.6	&	(	2.5	)	&	955.3	&	(	0.2	)	&	955.0	&	(	-0.1	)	&	955.1	&	(	0.0	)	&	954.8	&	(	-0.3	)	&	954.5	&	(	-0.6	)	\\
	&	1043.7	&	(	0.6	)	&	1043.7	&	(	0.6	)	&	1043.5	&	(	0.4	)	&	1043.1	&	(	0.0	)	&	1043.1	&	(	0.0	)	&	1042.5	&	(	-0.6	)	&	1041.9	&	(	-1.2	)	\\
	&	1226.9	&	(	1.2	)	&	1226.6	&	(	0.9	)	&	1226.4	&	(	0.6	)	&	1226.1	&	(	0.4	)	&	1225.8	&	(	0.0	)	&	1224.8	&	(	-1.0	)	&	1223.7	&	(	-2.1	)	\\
	&	1357.0	&	(	0.1	)	&	1357.1	&	(	0.1	)	&	1357.0	&	(	0.1	)	&	1356.8	&	(	-0.1	)	&	1356.9	&	(	0.0	)	&	1356.8	&	(	-0.1	)	&	1356.6	&	(	-0.3	)	\\
	&	1482.4	&	(	37.6	)	&	1444.5	&	(	-0.3	)	&	1442.9	&	(	-1.8	)	&	1442.6	&	(	-2.2	)	&	1444.8	&	(	0.0	)	&	1444.2	&	(	-0.6	)	&	1444.0	&	(	-0.8	)	\\
	&	1630.4	&	(	0.0	)	&	1629.6	&	(	-0.8	)	&	1630.4	&	(	0.0	)	&	1630.1	&	(	-0.2	)	&	1630.4	&	(	0.0	)	&	1630.2	&	(	-0.1	)	&	1630.1	&	(	-0.3	)	\\
	&	3288.6	&	(	277.8	)	&	3024.2	&	(	13.3	)	&	3013.8	&	(	2.9	)	&	3011.0	&	(	0.1	)	&	3010.8	&	(	0.0	)	&	3008.7	&	(	-2.2	)	&	3006.5	&	(	-4.4	)	\\
	&	3058.6	&	(	-0.2	)	&	3059.1	&	(	0.3	)	&	3059.0	&	(	0.2	)	&	3058.8	&	(	0.0	)	&	3058.8	&	(	0.0	)	&	3058.5	&	(	-0.3	)	&	3058.1	&	(	-0.7	)	\\
	&	3125.2	&	(	1.9	)	&	3125.3	&	(	2.0	)	&	3124.9	&	(	1.5	)	&	3124.2	&	(	0.9	)	&	3123.3	&	(	0.0	)	&	3121.0	&	(	-2.3	)	&	3117.9	&	(	-5.4	)	\\
	&	3148.4	&	(	2.2	)	&	3148.2	&	(	2.0	)	&	3147.8	&	(	1.5	)	&	3146.9	&	(	0.7	)	&	3146.3	&	(	0.0	)	&	3144.0	&	(	-2.2	)	&	3141.0	&	(	-5.2	)	\\
	&	\multicolumn{2}{|c|}{{}}	&	\multicolumn{2}{|c|}{{}}	&	\multicolumn{2}{|c|}{{}}	&	\multicolumn{2}{|c|}{{}}	&	\multicolumn{2}{|c|}{{}}	&	\multicolumn{2}{|c|}{{}}	&	\multicolumn{2}{|c|}{{}}			\\
MAX	&		&		277.8		&		&		13.3		&		&		6.6		&		&		2.7		&		&		0.0		&		&		5.6		&		&		12.3		\\
MAE	&		&		17.0		&		&		1.9		&		&		1.1		&		&		0.5		&		&		0.0		&		&		1.3		&		&		2.7		\\
RMSD	&		&		61.4		&		&		3.7		&		&		1.9		&		&		0.9		&		&		0.0		&		&		2.0		&		&		4.4		\\

\hline
\end{tabular}
}
\end{center}
\caption{Fundamental anharmonic frequencies obtained with iGVPT2 using different stepsizes in reduced coordinates. Using as reference the results obtained with a stepsize of 0.5, the max deviation (MAX), root-mean-square deviation (RMSD) and average unsigned deviation (MAD) values are given in the last rows. All values are given in cm$^{-1}$}
\label{table:reducedCoordinates}
\end{table}

\begin{table}[H]

\scriptsize{
\begin{center}
\setlength{\tabcolsep}{3pt}
\begin{tabular}{|r|rr|rr|rr|rr|rr|rr|r|}

\hline
Coord.	&	\multicolumn{2}{|c|}{{0.005}}	&	 \multicolumn{2}{|c|}{{0.010}}		&	 \multicolumn{2}{|c|}{{0.020}}		&	 \multicolumn{2}{|c|}{{0.025}}	&	 \multicolumn{2}{|c|}{{0.050}}		&	 \multicolumn{2}{|c|}{{0.075}}		\\
\hline
H$_2$O	&	1581.2	&	(	-0.7	)	&	1581.5	&	(	-0.4	)	&	1581.6	&	(	-0.3	)	&	1581.6	&	(	-0.3	)	&	1581.8	&	(	-0.1	)	&	1582.2	&	(	0.3	)		\\
	&	3680.0	&	(	1.1	)	&	3680.1	&	(	1.1	)	&	3679.9	&	(	1.0	)	&	3679.8	&	(	0.9	)	&	3678.6	&	(	-0.3	)	&	3677.1	&	(	-1.8	)		\\
	&	3777.1	&	(	5.2	)	&	3776.9	&	(	5.0	)	&	3776.1	&	(	4.2	)	&	3775.5	&	(	3.6	)	&	3770.3	&	(	-1.6	)	&	3761.6	&	(	-10.3	)		\\
H$_2$CO	&	1173.2	&	(	-0.1	)	&	1172.9	&	(	-0.4	)	&	1173.1	&	(	-0.3	)	&	1173.1	&	(	-0.3	)	&	1173.1	&	(	-0.2	)	&	1173.1	&	(	-0.2	)		\\
	&	1251.2	&	(	-1.5	)	&	1252.7	&	(	0.0	)	&	1252.8	&	(	0.1	)	&	1252.8	&	(	0.1	)	&	1252.7	&	(	0.0	)	&	1252.5	&	(	-0.2	)		\\
	&	1518.0	&	(	-0.1	)	&	1518.1	&	(	0.0	)	&	1517.9	&	(	-0.2	)	&	1518.3	&	(	0.1	)	&	1518.0	&	(	-0.1	)	&	1517.9	&	(	-0.3	)		\\
	&	1731.6	&	(	-0.7	)	&	1732.1	&	(	-0.2	)	&	1731.9	&	(	-0.4	)	&	1732.7	&	(	0.4	)	&	1731.9	&	(	-0.4	)	&	1732.0	&	(	-0.3	)		\\
	&	2825.4	&	(	0.8	)	&	2825.3	&	(	0.7	)	&	2825.2	&	(	0.6	)	&	2825.7	&	(	1.1	)	&	2824.7	&	(	0.0	)	&	2824.1	&	(	-0.6	)		\\
	&	2855.5	&	(	2.1	)	&	2856.4	&	(	2.9	)	&	2856.1	&	(	2.6	)	&	2856.0	&	(	2.6	)	&	2853.8	&	(	0.4	)	&	2850.4	&	(	-3.0	)		\\
C$_2$H$_4$	&	843.2	&	(	20.5	)	&	825.6	&	(	2.9	)	&	823.0	&	(	0.3	)	&	822.8	&	(	0.1	)	&	822.3	&	(	-0.4	)	&	821.6	&	(	-1.1	)		\\
	&	947.3	&	(	-0.4	)	&	955.4	&	(	7.7	)	&	931.2	&	(	-16.5	)	&	955.3	&	(	7.6	)	&	952.2	&	(	4.4	)	&	947.1	&	(	-0.6	)		\\
	&	957.0	&	(	1.9	)	&	965.5	&	(	10.4	)	&	955.4	&	(	0.3	)	&	963.0	&	(	7.9	)	&	954.9	&	(	-0.2	)	&	954.2	&	(	-0.9	)		\\
	&	1044.0	&	(	0.9	)	&	1043.5	&	(	0.4	)	&	1043.6	&	(	0.5	)	&	1043.6	&	(	0.5	)	&	1043.1	&	(	0.0	)	&	1042.3	&	(	-0.7	)		\\
	&	1235.6	&	(	9.8	)	&	1227.6	&	(	1.9	)	&	1226.7	&	(	1.0	)	&	1226.6	&	(	0.8	)	&	1226.1	&	(	0.3	)	&	1225.2	&	(	-0.5	)		\\
	&	1370.4	&	(	13.5	)	&	1358.1	&	(	1.2	)	&	1357.0	&	(	0.1	)	&	1357.0	&	(	0.1	)	&	1356.9	&	(	0.0	)	&	1356.7	&	(	-0.3	)		\\
	&	1451.0	&	(	6.3	)	&	1524.9	&	(	80.1	)	&	1445.9	&	(	1.2	)	&	1444.5	&	(	-0.3	)	&	1443.1	&	(	-1.7	)	&	1444.1	&	(	-0.7	)		\\
	&	1636.0	&	(	5.6	)	&	1632.1	&	(	1.7	)	&	1630.5	&	(	0.1	)	&	1630.9	&	(	0.6	)	&	1629.9	&	(	-0.4	)	&	1629.5	&	(	-0.9	)		\\
	&	1787.1	&	(	-1223.7	)	&	3333.8	&	(	323.0	)	&	3024.2	&	(	13.4	)	&	3018.4	&	(	7.6	)	&	3010.5	&	(	-0.4	)	&	3008.8	&	(	-2.0	)		\\
	&	3060.8	&	(	2.0	)	&	3058.8	&	(	0.0	)	&	3059.1	&	(	0.3	)	&	3059.1	&	(	0.3	)	&	3058.9	&	(	0.1	)	&	3058.7	&	(	-0.1	)		\\
	&	3126.8	&	(	3.5	)	&	3125.3	&	(	2.0	)	&	3125.4	&	(	2.1	)	&	3125.2	&	(	1.9	)	&	3123.9	&	(	0.6	)	&	3121.6	&	(	-1.7	)		\\
	&	3150.1	&	(	3.8	)	&	3148.6	&	(	2.3	)	&	3148.3	&	(	2.0	)	&	3148.1	&	(	1.9	)	&	3146.8	&	(	0.5	)	&	3144.5	&	(	-1.8	)		\\
	&	\multicolumn{2}{|c|}{{}}	&	 \multicolumn{2}{|c|}{{}}		&	 \multicolumn{2}{|c|}{{}}		&	 \multicolumn{2}{|c|}{{}}	&	 \multicolumn{2}{|c|}{{}}		&	 \multicolumn{2}{|c|}{{}}		\\
MAX	&		&		1223.7		&		&		323.0		&		&		16.5		&		&		7.9		&		&		4.4		&		&		10.3			\\
MAE	&		&		62.1		&		&		21.2		&		&		2.3		&		&		1.8		&		&		0.6		&		&		1.4			\\
RMSD	&		&		267.1		&		&		72.7		&		&		4.8		&		&		3.1		&		&		1.1		&		&		2.5			\\

\hline

\end{tabular}
\end{center}
}
\caption{Fundamental anharmonic frequencies obtained with iGVPT2 using same stepsize for all normal coordinates. Using as reference the results obtained with a stepsize of 0.5 with reduced coordinates, the max deviation (MAX), root-mean-square deviation (RMSD) and average unsigned deviation (MAD) values are given in the last rows. All values are given in cm$^{-1}$}
\label{table:normalCoordinates}
\end{table}


\subsection*{\sffamily \large Timings}
For large-sized molecules, the main bottleneck lies in the computation time of the third and fourth derivatives when they are calculated using an \emph{ab initio} or a DFT potential. 
Using iGVPT2, the real time of this part of calculation can be divided by $\approx f^3$ (where $f$ is the number of harmonic modes) provided access to a computer of $\approx f^3$ cores.  
As an example, we calculated the fundamental frequencies for a large molecule (the protonated dipeptide GlyGlyH+) using GVPT2 method and B3LYP/def2-TZVPP potential obtained via Orca\cite{orca} software. The total CPU time needed for this calculation is about 109888 hours. Using 2000 cores (of three computer centers) we were able to perform this calculation in only 56 (real time) hours.
The user may further reduce calculation time by using hybrid approaches implemented in iGVPT2. Using the B3LYP/def2-TZVPP method to compute harmonic modes and the molecular mechanic potential MMFF94, we were able to compute the anharmonic frequencies in only 0.5 hour (6 seconds for third and fourth derivatives). In one of our recent papers, we demonstrated that this hybrid method can reproduce the experimental fundamental anharmonic frequencies with a very good accuracy\cite{Barnes16}.

\section*{\sffamily \Large Conclusions}

In  this  paper, we have  presented a software "iGVPT2"  which  is designed  to  compute the anharmonic corrections to vibrational frequencies using the
VPT2 approach or its variants VPT2+K, DCPT2, HDCPT2. iGVPT2 supports several computation chemistry packages. Other CCP can be supported via a plugin (script or program) to be written by the user. 
Only energies and dipoles are needed. Any method implemented in the supported CCP can be used, including the methods where the analytical gradient is not yet implemented.
The real time calculation can be reduced due to our highly parallelized code. It can be also reduced using the hybrid approach implemented in iGVPT2. 
By its time-effectiveness and its accuracy, the hybrid DFT//MMFF94 implemented in iGVPT2 is a very interesting alternative to full DFT calculation where all derivatives are calculated at DFT level.  
iGVPT2 is available free of charge for non-commercial use. A user manual, plugins and several examples, one can find on the page \url{https://sites.google.com/site/allouchear/igvpt2}

\subsection*{\sffamily \large Acknowledgments}
This work was granted access to the HPC resources of the FLMSN, "F\'ed\'eration Lyonnaise de Mod\'elisation et Sciences Num\'eriques", partner of EQUIPEX EQUIP@MESO,
and to the "Centre de calcul CC-IN2P3" at Villeurbanne, France .
The authors are members of the Glycophysics Network (\url{http://glyms.univ-lyon1.fr})




\bibliographystyle{plain}
\bibliography{mybib}

\end{document}